\documentclass[12pt]{article}
\usepackage[centertags]{amsmath}
\usepackage{amssymb,amsfonts}
\usepackage[comma,super]{natbib}
\usepackage{latexsym}
\usepackage{graphicx}
\DeclareMathOperator{\sech}{sech}
\begin{document}
\begin{center}
{\Large \bf The effect of a two-fluid atmosphere on relativistic stars} \\
\vspace{1.5cm} {\bf Gabriel Govender, \bf Byron P. Brassel and \bf Sunil. D. Maharaj\footnote{email:maharaj@ukzn.ac.za
}
}\\
Astrophysics and Cosmology Research Unit,\\
School of Mathematics, Statistics and Computer Science,\\
University of KwaZulu-Natal,\\
Private Bag X54001,\\
Durban 4000,\\
South Africa\\
\vspace{1.5cm} {\bf Abstract}\\
\end{center}
We model the physical behaviour at the surface of a relativistic
radiating star in the strong gravity limit. The spacetime in the
interior is taken to be spherically symmetrical and shear-free. The heat conduction in the interior
of the star is governed by the geodesic motion of fluid particles
and a nonvanishing radially directed heat flux. The local atmosphere
in the exterior region is a two-component system
consisting of standard pressureless (null) radiation and an
additional null fluid with nonzero pressure and constant energy
density. We analyse the generalised junction condition for the
matter and gravitational variables on the stellar surface and
generate an exact solution. We investigate the effect of the exterior energy density on the
temporal evolution of the radiating fluid pressure, luminosty,
gravitational redshift and mass flow at the boundary of the star.
The influence of the density on the rate of gravitational collapse
is also probed and the strong, dominant and weak energy conditions
are also tested. We show that the presence of the additional null fluid has a significant effect on the dynamical evolution of the star.\\


\section{Introduction}

A compact star is formed when a massive star $(M_{*}\geq
8M_{\odot})$ breaks away from a state of hydrostatic equilibrium and
collapses under its own gravity. This situation usually arises when
the massive star has reached the end of the first phase of its
evolution. The stellar object that is formed at the end of the
gravitational collapse process is very dense and small in size with
a typical radius of the order of $(7-20) km$ for ultracompact
quark-gluon stars, strange stars and pulsars (including unmagnetized
neutron stars). As a result of the high mass density
in the resulting configuration, the gravitational field throughout
the interior is significantly strong and the star has effectively
transitioned into the so called strong gravity regime. At this point
the object is a far more enhanced self gravitating system and is
termed a relativistic star since the theory of general relativity
would be required if we are to construct a plausible and realistic
model for its global dynamics. For detailed and comprehensive
reviews on the theory of nonadiabatic gravitational collapse and
relativistic stars the reader is referred to the works of
Oppenheimer and Snyder \cite{opp&sny}, Penrose \cite{penrose},
Glendenning \cite{glen}, Strauman \cite{straum} and Shapiro and
Teukolsky \cite{shap&teu}, among others. During the contraction of
the massive star and the evolution of the compact object,
gravitational binding energy is converted into heat energy which is
used for ionization and
dissociation in the dense interior. The excess heat energy 
is dissipated 
as radiation across the stellar surface. This process is described
adequately for a relativistic star by constructing the junction
conditions at the surface. These equations relate the interior and
exterior matter variables as well as the respective geometries.\\

It was in the pioneering work of Oppenheimer and Snyder
\cite{opp&sny} that the significant role of general relativity in
modeling the nonadiabatic gravitational collapse of stellar bodies
was established. The dynamics was studied by considering the
contraction of a spherically symmetric dust cloud and it was
realised that relativistic gravity could adequately describe the
outflow of heat energy in the form of pressureless radiation which
is often referred to as null dust. While this seminal investigation
provided an initial framework in which to construct models for the
internal evolution of the dissipating object, it was not completely
clear what the conditions were at the boundary or surface. Moreover
it was also not clear as to what the appropriate geometry of the
exterior region was, i.e. there was no known exact solution to the
Einstein field equations that described the gravitational field of a
radiating star. Interestingly, it was some time later that the first
solution that describes this scenario was obtained, and this has
since then become a crucial and essential ingredient when studying
stellar like objects with heat flow and strong gravitational fields
in astrophysics. The Vaidya \cite{vaidya} metric is the most well
known and widely used solution in general relativity that describes
the exterior gravitational field of a dense object exhibiting
outward radial heat flow. It defines outgoing null radiation and is
written in terms of the mass of the radiating body. This result
provided a marked advancement in the modeling process and it was
eventually when the so called junction conditions for radiating
stars was generated by Santos \cite{santos} that general relativity
provided a satisfactory framework in which these objects could be
studied in greater detail. The Santos junction conditions are physically
very meaningful, for localised astrophysical
objects because they reveal that the pressure of the radiating
stellar fluid is not zero at the boundary but
rather that the pressure is proportional to the magnitude of the
heat flux. This seemingly simple mathematical relation has to
be solved as a highly nonlinear differential equation; the solutions
of which provide the evolution of the gravitational field potentials
and in turn the behaviour of the matter field. However, it must be
noted that this standard framework only describes the emission of
pressureless null radiation (photons) into the exterior region of
the dissipating relativistic star. It has not been used to probe the
outflow of any other type of observable radiation or elementary
particles like neutrinos, which are thought to be
significant carriers of heat energy in stars and released by
particle production processes at the stellar surface. Recently,
Maharaj \emph{et al} \cite{mgg} have generalised the Santos junction
condition by matching the interior geometry of a spacetime
containing a shear-free heat conducting stellar fluid to the
geometry of an exterior region that is described by the generalised
Vaidya metric which contains an additional Type \textrm{II} null
fluid. This new result provides a greater array of possibilities
with regards to the modeling of relativistic objects in astrophysics
as it depicts a more general exterior region for a radiating star,
which is comprised of a two-fluid system: a combination of the standard null
radiation as in the case of the Santos framework, and an additional
fluid distribution which is more general and can be taken to be
another form of radiation or better still, a field of particles such
as neutrinos, as mentioned above. An interesting feature of this
generalised junction condition is that the radiating fluid pressure at
the boundary is not only coupled to the internal heat flux but also
to the non-vanishing energy density of the Type \textrm{II} null
fluid. This conveys a direct relationship between the evolution of
the interior and exterior matter and gravitational fields, at the
boundary of the star and consequently may yield physical behaviour
that is far different and perhaps more realistic from that which
arises in the standard scenario.\\

A substantial amount of work on relativistic radiating stars has
been carried out in the standard Santos framework. In the
investigations of de Oliveira \emph{et al} \cite{deol} and Maharaj
and Govender \cite{m&g00}, the Santos junction conditions were
generalised to include the effects of an electromagnetic field and
shearing anisotropic stresses during dissipative stellar collapse.
More recently, analytical models for shear-free nonadiabatic
collapse in the presence of electric charge were obtained by
Pinheiro and Chan \cite{p&c}. The influence of pressure anisotropy,
shear and bulk viscosity on the density, mass, luminosity and
effective adiabatic index of an $8M_{\odot}$ contracting radiating
star was studied by Chan \cite{chan1, chan2}. These results were
later extended by Pinheiro and Chan \cite{p&c1}. Misthry \emph{et
al} \cite{mist} generated nonlinear exact models in the shear-free
regime for relativistic stars with heat flow, using a group
theoretic approach that involves the Lie symmetry analysis of the
Santos junction condition. Following this, Abebe \emph{et al}
\cite{gez}, employing the same technique, investigated the behaviour
of radiating stars in conformally flat spacetime manifolds. Abebe \emph{et al} \cite{ab, abe} used
Lie analysis to find radiating Euclidean stars with an equation of state. Collapse
models with internal pressure isotropy and vanishing Weyl stresses
were also probed by Maharaj and Govender \cite{m&g}. They
investigated the dynamical stability of the dissipating stellar
fluid and demonstrated that the configuration was more unstable
close to the centre than in the outer regions. It is also well
understood that the thermal evolution of the radiating fluid is
crucial in any stellar model and the precise role of the relaxation
and mean collision time were analysed by Martinez \cite{martinez},
Herrera and Santos \cite{her&san} and Govender \emph{et al}
\cite{govetal}. These ideas were further exploited in the more
recent works of Naidu \emph{et al} \cite{n1}, Naidu and Govender
\cite{n2} and Maharaj \emph{et al} \cite{mahetal}. Attempts to model
radiating stellar matter with a more realistic form, have been made
by imposing either a barotropic or polytropic equation of state for
the fluid distribution. Wagh \emph{et al} \cite{wagh} implemented a
linear equation of state for matter evolving in shear-free
spacetimes and Goswami and Joshi \cite{gos&josh} studied the
gravitational collapse of an isentropic perfect fluid with a linear
equation of state.\\

It is noteworthy, at this point, to mention that not much work has
been done in describing dissipating stellar bodies in the more
general scenario when the exterior region is a two-fluid system and
the Santos junction condition is generalised. Since the appearance
of the Maharaj \emph{et al} \cite{mgg} result in the literature, no
exact solution and/or corresponding physical  model has been
presented. However, it should be noted that the idea of a Type
\textrm{II} null fluid existing in the local exterior of a radiating
star has been explored previously in other contexts and without
there being any direct connection between the interior and exterior.
The physical properties of the generalised Vaidya spacetime
containing a Type \textrm{II} null fluid were studied extensively by
Husain \cite{husain}, for which exact solutions were obtained
strictly for the exterior fluid. In a subsequent investigation Wang
and Wu \cite{wangwu} further extended theses ideas and generated 
classes of solutions for spherically symmetric geometries. These
general results contain the well known monopole solution, the
de Sitter and anti-de Sitter solutions, the charged Vaidya models
and the radiating dyon solution. The conditions for physically
reasonable energy transport mechanisms in the generalised Vaidya
spacetime were explored by Krisch and Glass \cite{k&g05} and Yang
\cite{yang}. Glass and Krisch \cite{g&k98, g&k99} used the notion of
the null fluid to describe the qualitative features of a localised
string distribution in 4-dimensions, and generated analytical
solutions for null string fluids exhibiting pressure isotropy in the
limit of diffusion. This realisation that a Type \textrm{II} null
fluid can be used to describe a diffusing string on cosmological
scales, in particular, supports the earlier works of Vilenkin
\cite{vilenkin}, Percival \cite{perc} and Percival and Strunz
\cite{perc&str}.\\

This paper is organised as follows: In section 2 we present the
basic theory for relativistic radiating stellar models with
spherically symmetric shear-free interiors and two-fluid exteriors.
The generalised junction condition is then integrated as a nonlinear
second order differential equation at the boundary, and an exact
solution for constant null fluid energy density is generated. The
luminosities and gravitational redshift are then defined for the
case when the null fluid is present on the outside. Finally, some
qualitative features of the null fluid are briefly described. A
detailed physical analysis is carried out in section 3. The
evolution of the radiating fluid and the gravitational variables at
the surface, are probed and the standard and generalised models are
compared in order to establish the impact of the exterior null
fluid. This is achieved by constructing temporal profiles of the
radiating fluid pressure, the surface and asymptotic luminosities,
and the gravitational redshift using the exact solution due to
Thirukkanesh and Maharaj \cite{thiru&mah09} and the solution
presented in this work. The strong, dominant and weak energy
conditions, mass flow, and rate of gravitational collapse for the
two scenarios are finally investigated. In section 4 we discuss our
results and make concluding remarks.

\section{Constructing the model}
\subsection{The interior and exterior geometry}
The dynamics of the gravitational field in the stellar interior, in
the absence of shearing stresses, is governed by the spacetime line
element
\begin{equation}
ds^2 = -A^2(r, t)dt^2 + B^2(r, t)[dr^2 + r^2(d\theta^2 +
\sin^2{\theta}d\phi^2)],\label{metric}
\end{equation}
where $A(r, t)$ and $B(r, t)$ are the relativistic gravitational
potentials. The matter inside the star is defined by a
relativistic fluid with heat conduction in the form of the energy
momentum tensor
\begin{equation}
T_{ab}^{-} = (\mu + p)u_a u_b + pg_{ab} + q_a u_b + q_b
u_a.\label{inmat}
\end{equation}
Here $\mu$ and $p$ are the fluid energy density and isotropic
pressure respectively. The components $g_{ab}$ represent the metric tensor field and $ \bf u$
and $\bf q$, are the fluid four-velocity and the radial
heat flux, respectively. The Einstein field equations
$G_{ab}^-=T_{ab}^-$ for the interior may be written as
\begin{subequations}
\label{0sheareinfield}
\begin{eqnarray}
\mu &=& 3\frac{\dot{B}^2}{A^2B^2}-
\frac{1}{B^2}\left(2\frac{B^{\prime\prime}}{B}
-\frac{B^{\prime2}}{B^2}
+\frac{4}{r}\frac{B^{\prime}}{B}\right),\label{0sheareinfield1}\\
p &=& \frac{1}{A^2}\left(-2\frac{\ddot{B}}{B} -\frac{\dot{B}^2}{B^2}
+ 2\frac{\dot{A}}{A} \frac{\dot{B}}{B}\right) \nonumber\\
& & + \frac{1}{B^2}\left(\frac{B^{\prime 2}}{B^2} +
2\frac{A^{\prime}}{A}\frac{B^{\prime}}{B} +
\frac{2}{r}\frac{A^{\prime}}{A} +
\frac{2}{r}\frac{B^{\prime}}{B}\right),\label{0sheareinfield2}\\
p &=& -2\frac{\ddot{B}}{A^{2}B} +
2\frac{\dot{A}}{A^3}\frac{\dot{B}}{B} -
\frac{\dot{B}^2}{A^2B^2} + \frac{1}{r}\frac{A^{\prime}}{AB^2}\nonumber\\
& & + \frac{1}{r}\frac{B^{\prime}}{B^3} +
\frac{A^{\prime\prime}}{AB^2} - \frac{B^{\prime 2}}{B^4} +
\frac{B^{\prime\prime}}{B^3},\label{0sheareinfield3}\\
q &=& -\frac{2}{AB^2}\left(-\frac{\dot{B}^{\prime}}{B} +
\frac{B^{\prime}\dot{B}}{B^2} +
\frac{A^{\prime}}{A}\frac{\dot{B}}{B}\right),\label{0sheareinfield4}
\end{eqnarray}
\end{subequations}
where dots and primes denote differentiation with respect to
coordinate time $t$ and radial distance $r$ respectively.

In the local region outside the star the dynamics of the
gravitational field may be described by the generalised Vaidya outgoing
radiation metric which has the following form
\begin{equation}
ds^2 = - \left(1 - 2\frac{ m(v,\sf{r})}{\sf{r}}\right)dv^2 -2
dvd{\sf{r}} + {\sf{r^2}} (d\theta^2 +\sin^2\theta
d\phi^2),\label{p1genvadmet}
\end{equation}
where $m(v, \sf{r})$ represents the mass flow function at the
surface, and is related to the gravitational energy within a given
radius $\sf{r}$. The characteristic feature about the metric
$(\ref{p1genvadmet})$ is that the mass function also contains a spatial
dependence in the radial direction during dissipation; this is
significantly different form the standard scenario in which the mass
at the boundary only has a time dependance. Husain \cite{husain} and
Wang and Wu \cite{wangwu} have shown that an energy momentum tensor
that is consistent with $(\ref{p1genvadmet})$ is
\begin{equation}
T_{ab}^{+} = \varepsilon l_{a}l_{b} +
\left(\rho+P\right)\left(l_{a}n_{b}+l_{b}n_{a}\right)+Pg_{ab},\label{emt}
\end{equation}
which is a superposition of null radiation and an arbitrary null
fluid. Here $\varepsilon$ is the energy density of the photon radiation,
and $\rho$ and $P$ are the energy density and pressure of the
additional null fluid, respectively. In general, $T_{ab}^+$
represents a Type \textrm{II} fluid as defined by Hawking and Ellis
\cite{hawkellis}. This additional fluid component has also been
interpreted by Husain \cite{husain}, Wang and Wu \cite{wangwu},
Glass and Krisch \cite{g&k98,g&k99} and Krisch and Glass
\cite{k&g05} as a string fluid distribution in four dimensions. The
null vector $\bf l$ is a double null eigenvector of the energy
momentum tensor $(\ref{emt})$ and the vector $\bf n$ is normal to the
spatial hypersurface. For the spacetime metric $(\ref{p1genvadmet})$ and the energy
momentum tensor $(\ref{emt})$ the Einstein field equations for the
external local two-fluid stellar atmosphere can be written as
\begin{subequations} \label{exeinfield}
\begin{eqnarray}
\varepsilon &=& -2\frac{m_v}{\sf{r}^2},\label{exeinfield1}\\
\nonumber\\
\rho &=& 2\frac{m_{\sf{r}}}{\sf{r}^2},\label{exeinfield2a}\\
\nonumber\\
P &=& -\frac{m_{\sf{r}\sf{r}}}{\sf{r}}.\label{exeinfield3}
\end{eqnarray}
\end{subequations}

It is interesting to note that in the standard framework, the Santos
\cite{santos} junction condition tells us that on the boundary of
the star the pressure of the stellar fluid is proportional to the
magnitude of the heat flux
\begin{equation}
(p)_{\Sigma} = (qB)_{\Sigma}.\label{sjc}
\end{equation}
This condition has recently been extended to incorporate
the generalised Vaidya radiating solution. Maharaj \emph{et al}
\cite{mgg} have shown from first principles that the junction
condition (describing the dissipation) for the metric
$(\ref{p1genvadmet})$ can be written in the form
\begin{equation}
(p)_{\Sigma}=(qB-\rho)_{\Sigma}.\label{gjc}
\end{equation}
Note here, that this result is more general and contains the Santos
equation $(\ref{sjc})$ as a special case when $\rho=0$, i.e. when the
additional null fluid is absent and the outside contains only pure
radiation. It is also important to observe that the generalised
boundary equation $(\ref{gjc})$ is new for spherically symmetric
nonadiabatic stellar models and consequently have no reported exact
solutions. Furthermore, the new result is quite extensive and may be
applicable to more physically realistic astrophysical scenarios as
it includes the effect of the null fluid energy density. As
mentioned earlier, the null fluid and its nonvanishing energy
density have been explored in the context of cosmological and more
localised string fluid distributions in four dimensions. It
may also be appropriate for the description of neutrino
outflow from compact relativistic stellar objects in which
nonadiabatic and particle production processes prevail. In view of
this it is important to emphasize that the generalised null fluid
models should allow for significant improvement on the results
obtained by Glass \cite{glass}, following the seminal treatment by
Misner \cite{misner}.

\subsection{Qualitative features of the exterior null fluid}

For a better understanding of the impact of the additional null
fluid in the exterior of the radiating body, it is worthwhile to establish what the general features and
qualities of the fluid are. It is clearly evident from the exterior
field equations $(\ref{exeinfield2a})$ and  $(\ref{exeinfield3})$
that an explicit relationship between the energy density $\rho$ and
the pressure $P$ can be obtained. By simply differentiating through
$(\ref{exeinfield2a})$ with respect the exterior radial coordinate
$\sf{r}$ and using $(\ref{exeinfield3})$ we see that\\

\begin{equation}
P=-\frac{1}{2\sf{r}}\left(\sf{r}^2 \frac{d \rho}{d
\sf{r}}+2\sf{r}\rho\right).\label{presdensrelat}
\end{equation}
This equation indicates that there
exists, in general, a linear connection between the pressure and the
energy density. Moreover, $(\ref{presdensrelat})$ also suggests that
given a form for the density, we can determine both the magnitude
and the sign of the pressure. The sign of the pressure is crucial as it reveals more about the nature of the null
fluid and the interaction of the fluid with the gravitational field.
If an arbitrary functional form is prescribed for the density then it
is more difficult to determine the sign of the pressure, as this
would depend on the actual expressions for the terms in brackets.
However, if as in the case of the model presented in this work, the
density at the surface is taken to be constant then it is
straightforward to obtain the signature of $P$ since for physically
meaningful applications the energy density has to be strictly
positive $(\rho >0)$. With this in mind it turns out that for a
constant null fluid energy density at the surface, the pressure $P$
is always negative. This is consistent with the ideas exploited
in cosmological scenarios where a fluid characterised by the
cosmological constant or as dark energy or a scalar field, has to
have a negative pressure in order to counteract the attractive force
of gravity induced by matter and large scale structure in the
universe. In a more localised astrophysical setting as in this
study, the interpretation may be somewhat different.

\subsection{Luminosity and gravitational redshift}
In our new generalised framework, in which the exterior region of
the radiating body consists of both standard null radiation
(photons) and an additional Type \textrm{II} null fluid, we expect that the null fluid should have some effect on
the observable and measurable properties of the radiating fluid at
the surface. It is quite evident from the generalised junction
condition $(\ref{gjc})$ that the fluid pressure at the boundary now
depends on the energy density $\rho$ of the null fluid. Moreover, it
has also been well established by Bonnor \emph{et al} \cite{bonnor}
that the luminosity and redshift of a dissipating star in general
relativity depend on the fluid pressure; hence one expects the null
fluid to in some way influence these quantities at the surface.\\

We aim now to generalise the definitions for the fluid luminosity
and gravitational redshift in the context of our new model. The null
radiation emitted by a relativistic object with heat flow,
experiences a gravitational redshift (change in energy) due to its
motion in the gravitational field. This redshift is generally
written as

\begin{equation}
z_{\Sigma}=\sqrt{\frac{L_{\Sigma}}{L_{\infty}}}-1,\label{surfred}
\end{equation}
where $L_{\Sigma}$ is the luminosity of the radiating fluid at the
surface and $L_{\infty}$ is the luminosity of the radiating fluid as
seen by a stationary observer at an infinite distance away from the
object. These are given respectively, by
\begin{eqnarray}
L_{\Sigma} &=&
-\left(\frac{dv}{d\tau}\right)^2\frac{dm}{dv}\nonumber\\
\nonumber\\
&=& -\left(\frac{dv}{d\tau}\right)\frac{\partial m}{\partial
t}\frac{dt}{d\tau},\label{surflum}
\end{eqnarray}
and
\begin{eqnarray}
L_{\infty} &=& -\frac{dm}{dv}\nonumber\\
\nonumber\\
&=& -\frac{\partial m}{\partial
t}\frac{dt}{d\tau}\left(\frac{dv}{d\tau}\right)^{-1}.\label{linfinity}
\end{eqnarray}

\noindent In the above $m$ is the mass flow across the boundary $\Sigma$, $\tau$
is the timelike coordinate on the boundary, and $v$ is the timelike
coordinate in the exterior region described by the generalised
Vaidya spacetime $(\ref{p1genvadmet})$. (For details see
Maharaj \emph{et al} \cite{mgg}). The term $\frac{dv}{d\tau}$ is given in
terms of the interior gravitational potentials by

\begin{equation}
\frac{dv}{d\tau}=\left(r\frac{\dot{B}}{A}+\frac{(rB)^{\prime}}{B}
\right)^{-1}.\label{dvdtau}
\end{equation}

\noindent The derivative $\frac{dt}{d\tau}$ is written as

\begin{equation}
\frac{dt}{d\tau}=\frac{1}{A}.\label{dtdtau}
\end{equation}
The mass flow is given by
\begin{equation}
m(r,t)=\left[\frac{rB}{2}\left(1+r^2\frac{\dot{B}^2}{A^2}-\frac{1}{B^2}(B+rB')^2\right)\right]_\Sigma.\label{masssy}
\end{equation}
Differentiating (\ref{masssy}) results in the expression
\begin{equation}
\left(\frac{\partial m}{\partial
t}\right)_{\Sigma}=\left(\frac{r^3}{2}\frac{\dot{B}^3}{A^2}+\frac{r^3
B\dot{B}\ddot{B}}{A^2}-\frac{r^3 B\dot{B}^2 \dot{A}}{A^3}-r^2
\dot{B}^{\prime}-\frac{r^3
B^{\prime}\dot{B}^{\prime}}{B}+\frac{r^3}{2}\frac{B^{\prime
2}\dot{B}}{B^2}\right)_{\Sigma}.
\label{byron}
\end{equation}
Then using the generalised junction condition (\ref{gjc}) as well as (\ref{0sheareinfield2}) in (\ref{byron}) gives
\begin{equation}
\frac{\partial m}{\partial
t}=-\frac{1}{2}pr^2B^2\left(r\dot{B}+A+r\frac{AB^{\prime}}{B}\right)-\frac{1}{2}\rho
r^2AB^2\left(1+r\frac{B^{\prime}}{B}\right),\label{newstdder}
\end{equation}
which is the generalised result for the mass flow rate when the null fluid is present in the exterior with the generalised Vaidya metric with $m=m(v,r)$.
When the null fluid is absent $(\rho=0)$ we then get
\begin{equation}
\frac{\partial m}{\partial
t}=-\frac{1}{2}\left(pr^3B^2\dot{B}+qr^2AB^2(B+rB^{\prime})\right),\label{stdder}
\end{equation}
which corresponds to the mass flow rate when only null dust is present in the exterior with the standard Vaidya metric with $m=m(v)$. Hence we have shown that the mass flow rate is fundamentally affected by the 2-fluid atmosphere with the generalised Vaidya metric since the null fluid density $\rho$ is nonzero.

Using the derivatives
$(\ref{newstdder})$, $(\ref{dtdtau})$, and $(\ref{dvdtau})$ in the
definitions $(\ref{surflum})$ and $(\ref{linfinity})$ we generate
the following expressions for the surface luminosity $L_{\Sigma}$
and the asymptotic luminosity $L_{\infty}$, respectively
\begin{subequations}
\label{newlumins}
\begin{eqnarray}
L_{\Sigma} &=&
\frac{1}{2}r^2B^2\left[p\left(r\frac{\dot{B}}{A}+\frac{(rB)^{\prime}}{B}
\right)+\rho
\left(1+r\frac{B^{\prime}}{B}\right)\right]\left(r\frac{\dot{B}}{A}+\frac{(rB)^{\prime}}{B}
\right)^{-1},\label{newsurflum}\\
\nonumber\\
L_{\infty} &=&
\frac{1}{2}r^2B^2\left[p\left(r\frac{\dot{B}}{A}+\frac{(rB)^{\prime}}{B}
\right)+\rho
\left(1+r\frac{B^{\prime}}{B}\right)\right]\left(r\frac{\dot{B}}{A}+\frac{(rB)^{\prime}}{B}
\right).\label{newaslumin}
\end{eqnarray}
\end{subequations}
Equations (\ref{newsurflum}) and (\ref{newaslumin}) are the
generalisations of the surface and asymptotic luminosities when the
null fluid is present in the exterior region. It is clear
that the null fluid should have a marked impact on the values of
these quantities, as the energy density $\rho$ appears explicitly through the boundary pressure $p_{\Sigma}=(qB-\rho)_{\Sigma}$, and the gravitational
potentials $A$ and $B$, in the above forms. Furthermore $L_{\Sigma}$
and $L_{\infty}$ differ by a factor of
$\left(r\frac{\dot{B}}{A}+\frac{(rB)^{\prime}}{B} \right)^2$. We
observe, once again, that when the null fluid is absent in the
exterior $(\rho=0)$, $(\ref{newsurflum})$
and $(\ref{newaslumin})$ reduce to the following forms,

\begin{subequations}
\label{standardluminosities}
\begin{eqnarray}
L_{\Sigma}&=&\frac{1}{2}r^2B^2p,\label{stdsurflum}\\
\nonumber\\
L_{\infty}&=&\frac{1}{2}r^2B^2p\left(r\frac{\dot{B}}{A}+\frac{(rB)^{\prime}}{B}
\right)^2,\label{stdalum}
\end{eqnarray}
\end{subequations}
which are just the classical definitions for the luminosities when
there is only radiation in the outside of the dissipating body. We emphasise the point that
the luminosities $L_{\Sigma}$ and $L_{\infty}$ are fundamentally charged by the
2-fluid atmosphere with the generalised Vaidya metric with $m=m(v,r)$ since the null fluid energy density affects the spacetime geometry.
Investigating the spatial and temporal behaviour of $L_{\Sigma}$ and
$L_{\infty}$ is of crucial importance when modeling dense stars in
astrophysics. For a particular model of a spherically symmetrical
relativistic star with heat flow, Tewari \cite{tewari} used his
exact solution at the stellar boundary to compute $L_{\Sigma}$ and
$L_{\infty}$. He demonstrated that the luminosity at infinity could
be as much as $2.61089\times 10^{48}$ ergs/sec and that the surface
luminosity was of the order of $9.8\times 10^{48}$ ergs/sec. Sarwe
and Tikekar \cite{s&t} analysed the evolution of the asymptotic
luminosity and used it to determine the timescale for which a
$3.24M_{\bigodot}$ compact star with radius in the order of $17 km$
collapses to form a black hole.\\

Utilising the system $(\ref{newlumins})$ and the definition
$(\ref{surfred})$ we may write down the form for the gravitational
redshift of the radiating fluid at the surface as
\begin{equation}
z_{\Sigma}=\left(r\frac{\dot{B}}{A}+\frac{(rB)^{\prime}}{B}\right)^{-1}-1.\label{newsurfred}
\end{equation}
Although it appears that the structural form of the redshift has not
changed, the magnitude of $z_{\Sigma}$ will be
different since the potentials $A$ and $B$ are obtained by solving
equation $(\ref{gjc})$ (which when written in expanded form is a
second order nonlinear differential equation that describes the
generalised two-fluid system in the exterior) at the boundary.
Computing the gravitational redshift is useful when modeling
relativistic objects with heat flow as the form and the magnitude of
$z_{\Sigma}$ can be used to give a measure of compactness or denseness of the object; that is to say, measuring the redshift
determines the compactness of the relativistic body. Hladik and
Stuchlik \cite{h&s} have studied the gravitational redshift of
photons and neutrinos radiated by neutron stars and quark stars in
the braneworld scenario. They pointed out that when observational
data for the photon and neutrino surface redshift is known, their
model can be used to determine the critical parameters that describe
the compactness of the star under investigation.

\subsection{Generating an exact solution}
We now turn our attention to the process of generating an exact
solution to the new boundary condition $(\ref{gjc})$. Considering the high degree of nonlinearity and other
mathematical complexities that surround equation $(\ref{gjc})$ it is
convenient at this point to focus on relativistic stellar models in
which the fluid particles move along null geodesics from the core
and up through to the stellar surface where the radiation is lost to
the exterior. Such models were investigated by Rajah and Maharaj
\cite{raj&mah08}, Govender and Thirukkanesh \cite{gov&thiru09} and
Thirukkanesh and Maharaj \cite{thiru&mah09}. In particular, the
Govender and Thirukkanesh \cite{gov&thiru09} model investigated the
behaviour of a radiating star by including a cosmological constant
term in the Einstein field equations for the interior. It must
be pointed out, however, that although their resulting junction
condition has the same structure as in this study,
their model only accounts for the presence of pure null radiation in
the exterior. The motivation for this idea is that the dissipating
star may be immersed in an ambient environment that contains dark
energy which is not connected in any way to the interior of the
body. We do not follow the same line of thought but treat the null
fluid distribution in the most general way, and regard it as more
likely to be sourced from within the star during the radiating
phase. Consequently the fluid may be a form of high energy radiation
or a field of elementary particles such as neutrinos or even
electron-positron pairs as in the case of strongly magnetized
neutron stars. The studies mentioned above, were carried out in the
framework of the standard Santos junction condition $p=qB$. It is
our aim now to extend these models by utilising the new generalised
junction condition $(\ref{gjc})$, and make use of the fact that
the stellar atmosphere is now a well defined two-fluid system. For
our investigations, we consider the null fluid energy density $\rho$
to be constant on the stellar surface. This assumption is not
physically unreasonable as this situation corresponds to the
diffusion of the null fluid in the stellar exterior as the
relativistic star approaches a state of hydrostatic equilibrium and
enters the early stages of thermal cooling.

For geodesic motion of fluid particles we have the
gravitational potential $A=1$. The condition of pressure isotropy in the absence of
shearing fluid stresses admits the following analytical form
\begin{equation}
B(r, t) = -\frac{d}{c_{1}(t)r^2-c_{2}(t)},\label{pissol}
\end{equation}
for the gravitational potential $B$. It is also important to recall
that the above form corresponds to conformally flat tidal
gravitational effects and was first obtained by Banerjee \emph{et
al} \cite{banner}. Here $d$ is an arbitrary constant and $c_{1}(t)$
and $c_{2}(t)$ are functions of integration which have to be
determined in order to complete the exact solution. The fluid pressure and radial heat
flux reduce to the following equations
\begin{subequations}
\label{geodesiceqns}
\begin{eqnarray}
p &=&
\left(-2\frac{\ddot{B}}{B}-\frac{\dot{B}^2}{B^2}\right)+\frac{1}{B^2}\left(\frac{B^{\prime
2}}{B^2}+\frac{2}{r}
\frac{B^{\prime}}{B}\right),\label{geodesicfluidpressure}\\
\nonumber\\
q &=&
-\frac{2}{B^2}\left(-\frac{\dot{B}^{\prime}}{B}+\frac{\dot{B}B^{\prime}}{B^2}
\right),\label{geodesicheatflux}
\end{eqnarray}
\end{subequations}
from (\ref{0sheareinfield2}) and (\ref{0sheareinfield4})
respectively. With the above
system $(\ref{geodesiceqns})$ and the form $(\ref{pissol})$ the
generalised junction condition given by $(\ref{gjc})$ can be fully
expanded as
\begin{equation}
-4db(\dot{c_{1}}c_{2}-c_{1}\dot{c_{2}})(c_{1}b^{2}-c_{2})-4c_{1}c_{2}(c_{1}b^{2}-c_{2})^{2}
-2d^{2}(\ddot{c_{1}}b^{2}-\ddot{c_{2}})(c_{1}b^{2}-c_{2})\nonumber
\end{equation}
\begin{equation}
+5d^{2}(\dot{c_{1}}b^{2}-\dot{c_{2}})^{2} -\rho
d^{2}(c_{1}b^{2}-c_{2})^{2}=0,\label{hardform}
\end{equation}
where we have taken $r=r_{\Sigma}=b(=$ constant), on the stellar
surface. The value $r_{\Sigma}=b$ is the actual radius of the stellar
distribution and for compact and ultracompact relativistic bodies
like neutron stars and pulsars, and strange or quark stars, is of
the order of $7-30kms$. It is important to note the appearance of
the additional term that arises due to the presence of the null
fluid energy density $\rho \neq 0$. This parameter is taken to be
constant on the stellar boundary and as mentioned earlier, is an
appropriate and reasonable assumption. In the limit when the density
$\rho$ goes to zero, (for which the stellar exterior is composed
only of pure radiation), we regain the corresponding equation
obtained by Thirukkanesh and Maharaj \cite{thiru&mah09} in the
context of the standard framework. That equation has been studied
thoroughly and one of the earliest known solutions obtained was presented by Kolassis \emph{et al} \cite{kolassis}.
Our new equation $(\ref{hardform})$ now has an added degree of
complexity as a consequence of the null fluid component and
analysing it is more involved than in previous cases.

We realise that in order to integrate $(\ref{hardform})$, the following
transformation can be utilised
\begin{equation}
u(t)=c_{1}b^2-c_{2}.\label{labelchoice}
\end{equation}
Then equation $(\ref{hardform})$ may be rewritten as
\begin{equation}
4bdu^{2}\dot{c_{1}}+4(u^2-bd\dot{u})uc_{1}-4b^2u^2c_{1}^2=d^2\left[(2u\ddot{u}-5\dot{u}^2)+\rho
u^2\right].\label{riccatimasterequation}
\end{equation}
Equation $(\ref{riccatimasterequation})$ is a Riccati equation in
$c_{1}$ but is still difficult to solve in general. If we let
$u=\alpha$ (constant) then $(\ref{riccatimasterequation})$ becomes
\begin{equation}
\dot{c_{1}}+\frac{\alpha}{bd}c_{1}-\frac{b}{d}c_{1}^2=\frac{d}{4b}\rho.
\label{littleomega}
\end{equation}
We now make use of the transformation
\begin{equation}
c_{1}=-\frac{d}{b}\frac{\dot{w}}{w},\label{transformation}
\end{equation}
where $w(t)$ is an arbitrary function. Then equation
$(\ref{littleomega})$ becomes
\begin{equation}
\ddot{w}+\frac{\alpha}{bd}\dot{w}+\frac{\rho}{4}w=0,\label{bigomega}
\end{equation}
which is a second order linear ordinary differential equation with
constant coefficients. The general
solution to equation $(\ref{bigomega})$ is given by
\begin{equation}
w(t)=g_{1}(t)\exp\left[\lambda_{1}t\right] +g_{2}(t)\exp
\left[\lambda_{2}t\right].\label{solforw}
\end{equation}
In the above solution $g_{1}(t)$ and $g_{2}(t)$ are functions of
integration and
\begin{equation}
\lambda_{1}=\frac{1}{2}\left(\sqrt{\frac{\alpha
^2}{b^2d^2}-\rho}-\frac{\alpha}{bd}\right),\quad\quad\quad\quad
\lambda_{2}=-\frac{1}{2}\left(\sqrt{\frac{\alpha
^2}{b^2d^2}-\rho}+\frac{\alpha}{bd}\right).\label{roots}
\end{equation}
Then the functions $c_{1}(t)$ and $c_{2}(t)$ become
\begin{subequations}
\label{cforms}
\begin{eqnarray}
c_{1}(t) &=&
\left(-\frac{d}{b}\right)\left[\frac{g_{1}\lambda_{1}\exp(\lambda_{1}t)+g_{2}\lambda_{2}\exp(\lambda_{2}t)}
{g_{1}\exp(\lambda_{1}t)+g_{2}\exp(\lambda_{2}t)}\right],\label{formforc1}\\
\nonumber\\
c_{2}(t) &=&
-(bd)\left[\frac{g_{1}\lambda_{1}\exp(\lambda_{1}t)+g_{2}\lambda_{2}\exp(\lambda_{2}t)}
{g_{1}\exp(\lambda_{1}t)+g_{2}\exp(\lambda_{2}t)}\right]-\alpha.\label{formforc2}
\end{eqnarray}
\end{subequations}
Consequently the gravitational potential $B$ has the form
\begin{equation}
B(r,
t)=\frac{-db}{d\left[\frac{g_{1}\lambda_{1}\exp(\lambda_{1}t)+g_{2}\lambda_{2}\exp(\lambda_{2}t)}
{g_{1}\exp(\lambda_{1}t)+g_{2}\exp(\lambda_{2}t)}\right](b^2-r^2)+b\alpha}.\label{newbform}
\end{equation}
and the exact solution in metric form is
\begin{equation}
ds^2=-dt^2+\frac{d^2b^2}{\left[{d\left[\frac{g_{1}\lambda_{1}\exp(\lambda_{1}t)+g_{2}\lambda_{2}\exp(\lambda_{2}t)}
{g_{1}\exp(\lambda_{1}t)+g_{2}\exp(\lambda_{2}t)}\right](b^2-r^2)+b\alpha}\right]^2}\left[dr^2+r^2\left(d\theta
^2+\sin^2\theta d\phi^2\right)\right].\label{exactsolution}
\end{equation}
The new exact solution $(\ref{exactsolution})$ is similar, in
structure, to the solution found by Govender and Thirukkanesh
\cite{gov&thiru09}. However our model results from a different
physical scenario since the atmosphere of our star does not contain
the cosmological constant, but comprises a two-fluid system in which
one of the components is a string fluid. This model corresponds to
geodesic heat dissipation in a relativistic star when the string
fluid in the stellar atmosphere is undergoing diffusion. The
solution $(\ref{exactsolution})$ can now be used in the framework of
irreversible causal and noncausal thermodynamics to study the temperature evolution of the radiating star.

\newpage

\section{Physical analysis}
In order to establish the role and influence of the null fluid
energy density $\rho$ at the surface $\Sigma$ of the radiating
star we probe the temporal behaviour of the fluid and gravitational
variables in the generalised scenario with $m=m(v,r)$ described in this instance by
the exact solution $(\ref{exactsolution})$. We compare this with the standard case with $m=m(v)$
when $\rho=0$, using the solution which can be found in Thirukkanesh
and Maharaj \cite{thiru&mah09}. We point out that these models are
cast in natural units and consequently all physical parameters and
mathematical constants are dimensionless.

For the purpose of this investigation we plot the quantities $B$, $p$, $z$, $L_{\Sigma}$, $L_{\infty}$, $m$ and
$\frac{dm}{dt}$ in Fig. 1-12. We show that the null density $\rho$ drastically affects the physical behaviour of the model in the graphical plots.
For the purpose of the graphs we make the following choice for the
constants that appear in the exact solution:
\begin{center}
$\alpha=1$, $b=10$, $d=-1$, $g_1=g_2=1$.
\end{center}
For these values, $(\ref{roots})$ places a constraint on the
allowed values for the energy density $\rho$ and it becomes clear
that $0<\rho<1$. This restriction on $\rho$ is unique for this model
and our particular choice of values; it is therefore model dependant
and will differ from another exact solution.

\subsection{Temporal evolution of the potential $B$}

The behaviour of the gravitational potentials $B_{\rho=0}$ and $B_{\rho\neq0}$ is given in Fig. 1.
It is evident that the potential $B$ is
always significantly larger in the case when $\rho\neq0$ than in the
limiting scenario when $\rho=0$. In the former, $B$ is regular and
decreases steadily with time, and increases in the latter. This
trend suggests that the exterior null fluid has a suppressing effect
on the gravitational field at the surface. We observe that
this may in part be due to the nature and structure of our exact
solution, and can possibly be somewhat different
in another model.

\begin{figure}[t]
\centering
\includegraphics[scale=1.2]{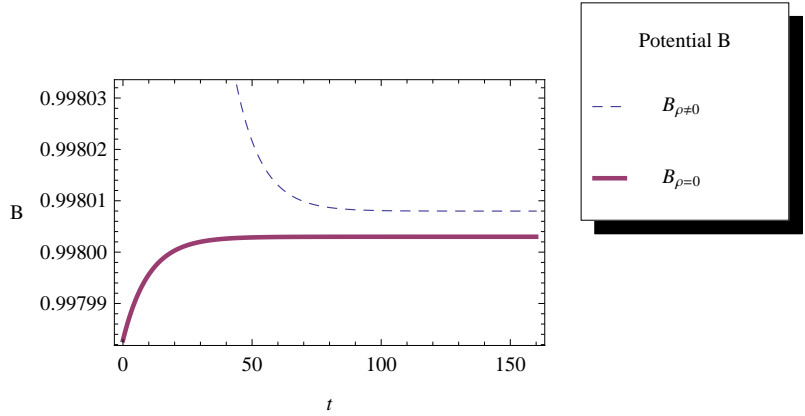} \caption{Temporal evolution of the gravitational potential $B$ at the stellar surface} \label{fig1}
\end{figure}

\newpage

\subsection{Temporal evolution of the fluid pressure $p_\Sigma$}
Here we examine the temporal behavior of the radiating fluid
pressure at the boundary of the object. The pressure, for the two
cases, is defined by the junction condition $(\ref{gjc})$. Fig. 2
shows that for the case of pressureless radiation only, the fluid
pressure $p_s$, at the boundary is very small $(p\approx 4\times
10^{-4})$ and decreases monotonically with time until it is negligible. This is consistent with what has been
theoretically established for stellar models in both the Newtonian
and relativistic limits. On the other hand, in Figure 3 it can be
seen that when $\rho\neq0$, $p$ is slightly suppressed and
becomes negative at the boundary. This is not surprising
considering the general form of $(\ref{gjc})$. Another reason
that supports this result is that when $\rho=0$, the pressure is
already close to zero, and if the density which is always strictly
positive (however small in magnitude it may be) is taken into
account it is clear that this will induce a further reduction. We
have also investigated the effect of varying the magnitude of $\rho$
and the family of profiles in Fig. 3 reveal that as the density is
increased within the the allowed range, the pressure becomes
increasingly more negative. This trend suggests that $p$ scales
inversely with $\rho$ at the surface
of the radiating star. It should be emphasized that despite this
behaviour, the pressure is still only just slightly negative and
does not in any way make the model physically unreasonable.

\begin{figure}[t]
\centering
\includegraphics[scale=1.0]{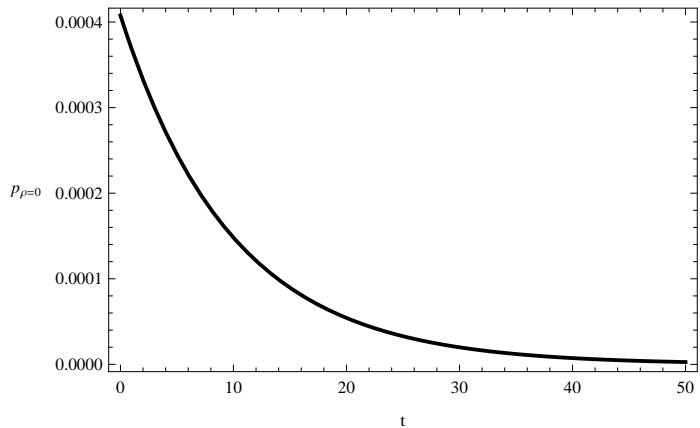} \caption{Temporal evolution of the radiating fluid pressure in the absence of $\rho$} \label{fig2a}
\end{figure}

\begin{figure}[t]
\centering
\includegraphics[scale=1.15]{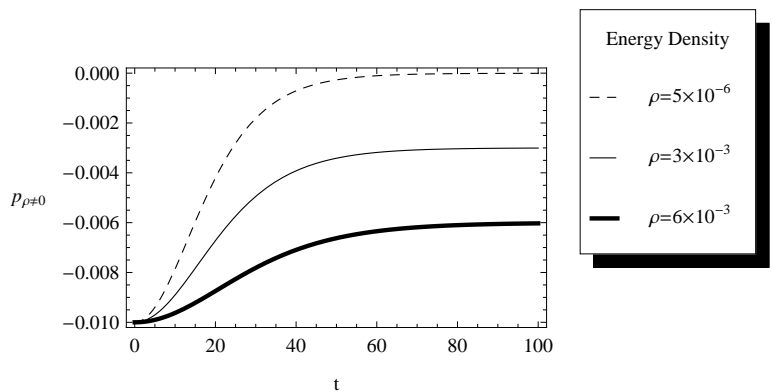} \caption{Temporal evolution of the radiating fluid pressure in the presence of $\rho$ with varying magnitudes} \label{fig2b}
\end{figure}

\newpage

\subsection{Temporal evolution of the fluid luminosity $L_\Sigma$}

Utilising the definitions (\ref{newsurflum}) and
(\ref{newaslumin}) along with the solution $(\ref{exactsolution})$
we are able to generate numerical profiles for the luminosity of
the radiating fluid at the surface of the relativistic body and the
luminosity as seen by a distant observer. In Fig. 4 we find that $L_{\Sigma}$ evolves in an acceptable manner and
decreases through positive values until it eventually becomes
substantially low. The plot also indicates that the initial
magnitude of the luminosity at the boundary is very small
$(L_{\Sigma}\approx 2.0\times 10^{-2})$, which may be justifiable
for stellar objects that are ultra-dense and possibly in the post
neutron star phase of its evolution. After an extended but finite
period of time the star becomes virtually non-luminous and may not
be observable. This is certainly consistent with the fact that the
so called quark and quark-gluon stars have not yet been detected
with current observing instruments. The family of
profiles in Fig. 5 exhibits very interesting behaviour for
$L_{\Sigma}$ in the more general situation when $\rho$ is
non-vanishing. In general, the luminosity is suppressed and changes
through negative values as the magnitude of $\rho$ is
increased; $L_{\Sigma}$ at early times is less negative, and at late
times tends to zero. The overall behaviour is similar to that
found in Fig. 4 and again may be applicable to stars in the strong
gravity regime. In Fig. 7, when there is only radiation in the exterior, the luminosity at infinity
is singular closer to the centre of the fluid distribution and evolves through positive but small values
with $r$. It is also clear that for large $r$ $(r\rightarrow\infty)$, $L_{\infty}$ tends to zero.
This suggests that a horizon will form at the end of the collapse of the radiating object. Fig. 6 on the other hand
indicates that $L_{\infty}$, when $\rho\neq0$, is finite at $r=0$ and evolves through negative values until eventually going to zero.
Again, this behaviour demonstrates the formation of a horizon at late times.

\begin{figure}[t]
\centering
\includegraphics[scale=1.0]{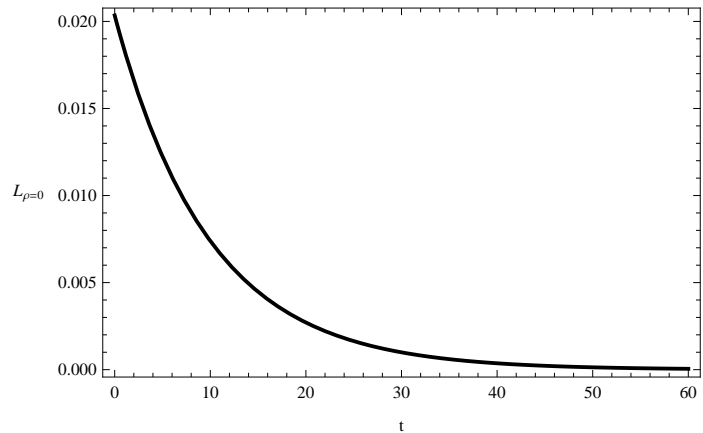} \caption{Temporal evolution of the radiating fluid luminosity in the absence of $\rho$} \label{fig3a}
\end{figure}

\begin{figure}[t]
\centering
\includegraphics[scale=1.1]{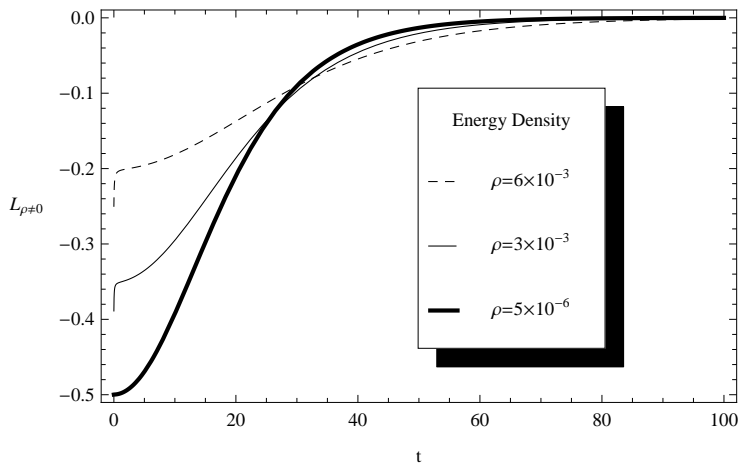} \caption{Temporal evolution of the radiating fluid luminosity in the presence of $\rho$ with varying magnitudes} \label{fig3b}
\end{figure}

\begin{figure}[t]
\centering
\includegraphics[scale=1.1]{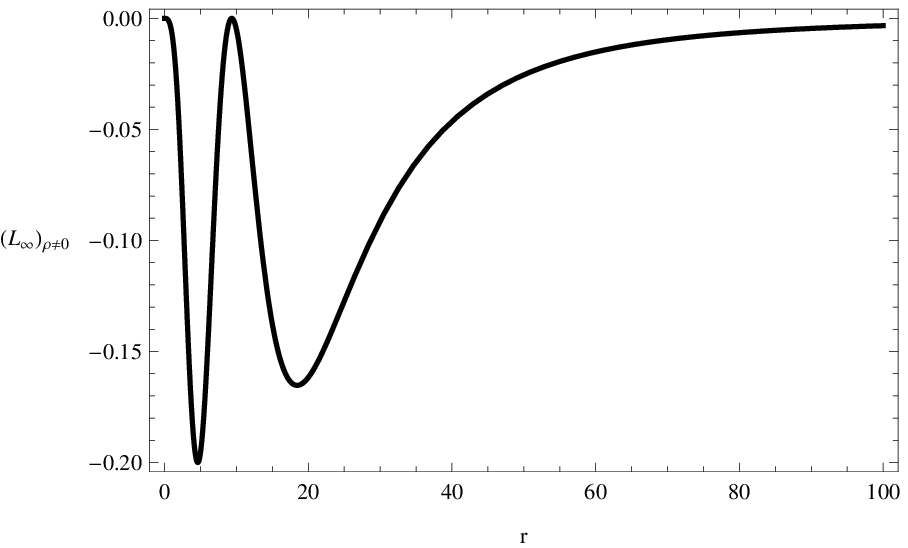} \caption{Temporal profile of the radiating fluid luminosity at infinity in the presence of $\rho$} \label{fig3b}
\end{figure}

\begin{figure}[t]
\centering
\includegraphics[scale=1.1]{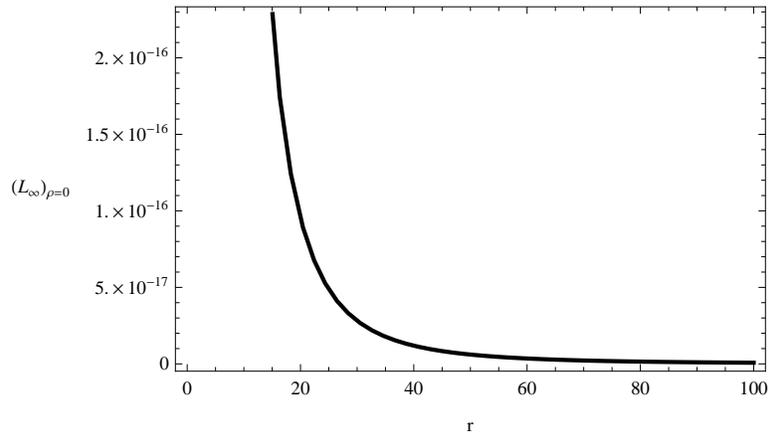} \caption{Temporal profile of the radiating fluid luminosity at infinity in the absence of $\rho$} \label{fig3b}
\end{figure}

\newpage

\subsection{Temporal evolution of the gravitational redshift $Z_\Sigma$}

The plots in Fig. 8 show that $Z_{\rho\neq0}>Z_{\rho=0}$ always. Furthermore,
it is clear that the redshift, in both the standard Vaidya and generalised Vaidya
models, evolves continuously and regularly through negative values.
This is probably largely due to the nature of the exact solutions.

\begin{figure}[t]
\centering
\includegraphics[scale=1.2]{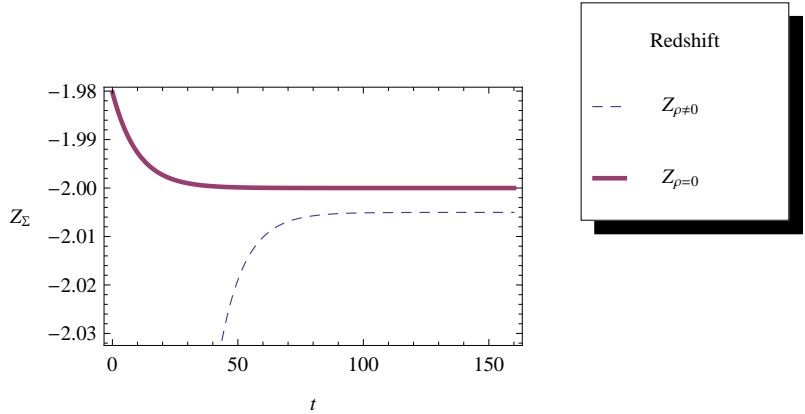} \caption{Temporal evolution of the gravitational redshift at the stellar surface} \label{fig4}
\end{figure}

\newpage

\subsection{Mass flow across the boundary}

Here we investigate the temporal behaviour of the mass flow and rate
of mass flow at the radiating surface. The numerical profile in
Fig. 9 illustrates that in the case of null dust $(\rho=0)$,
the mass flow across the radiating surface decreases steadily with
time, through positive values, until it terminates completely at
later times. The situation when the flow of mass stops altogether
corresponds to when the stellar object departs from its radiative
phase and transits into an equilibrium state, usually towards the
very late stages of gravitational collapse. For the model where the
additional null fluid $(\rho\neq0)$ is present in the exterior, Fig. 10
indicates that $m$ increases steadily with time and
eventually converges to a non-zero limiting value. At this point the
star is still losing mass but at a uniform rate. Hence, we may
conclude that in the standard Vaidya model the mass flow naturally comes to
a stop. However, in the more realistic and generalised Vaidya case, mass
continues to flow across the boundary for an extended period of
time. This is strong motivation for the idea that the null fluid
distribution in the local exterior is actually sourced from within
the radiating body and is released along with radiation during
dissipation. This must then mean that the null fluid is more likely
to be a distribution of particles like neutrinos which is consistent
with what is believed to be the case for most compact objects (see the earlier treatments of Misner \cite{misner} and Glass
\cite{glass}). In Fig. 11 we observe that mass flow rate across the
surface, decreases fairly rapidly and ultimately goes to zero when
the flow ceases. The profile in Fig. 12 indicates that for
$\rho\neq0$ $(dm/dt)$ is always negative and would seem to
contradict the profile in Fig. 10. This is probably due to
the form of the gravitational potential in the solution
$(\ref{exactsolution})$ and the choice of parameter values that have
been considered.

\begin{figure}[t]
\centering
\includegraphics[scale=1.0]{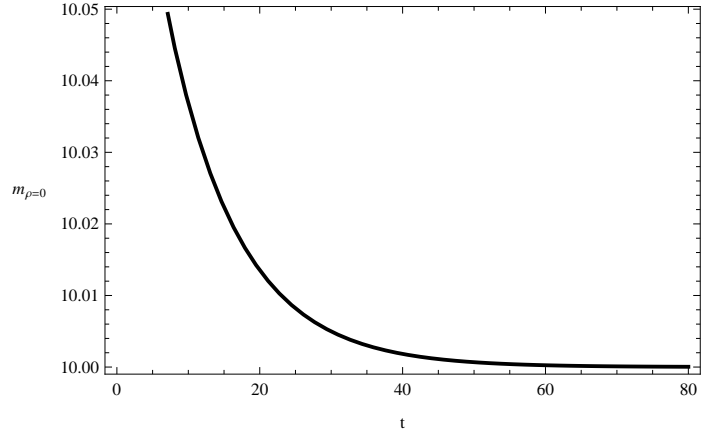} \caption{Temporal evolution of the mass flow across the stellar surface, in the absence of $\rho$} \label{fig4}
\end{figure}

\begin{figure}[t]
\centering
\includegraphics[scale=1.0]{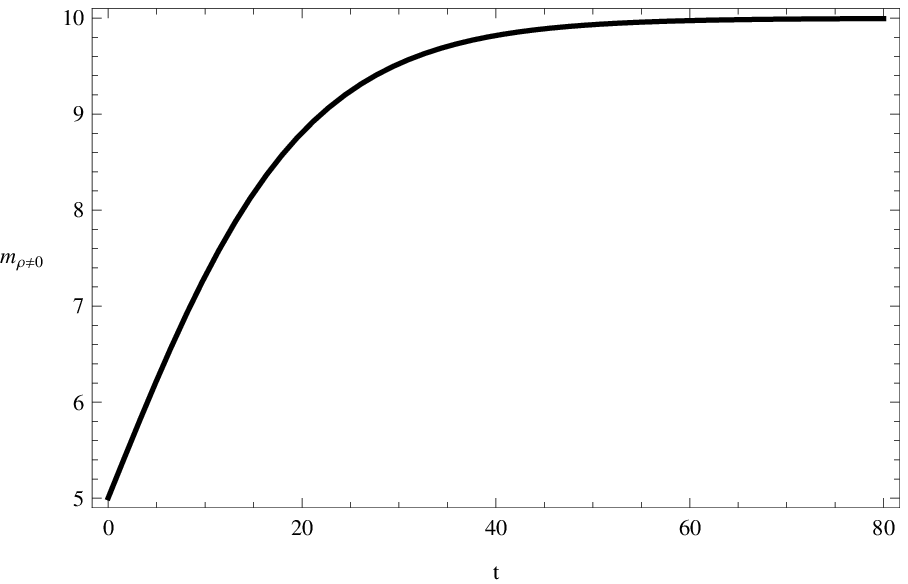} \caption{Temporal evolution of the mass flow across the stellar surface, in the presence of $\rho$} \label{fig4}
\end{figure}

\begin{figure}[t]
\centering
\includegraphics[scale=1.0]{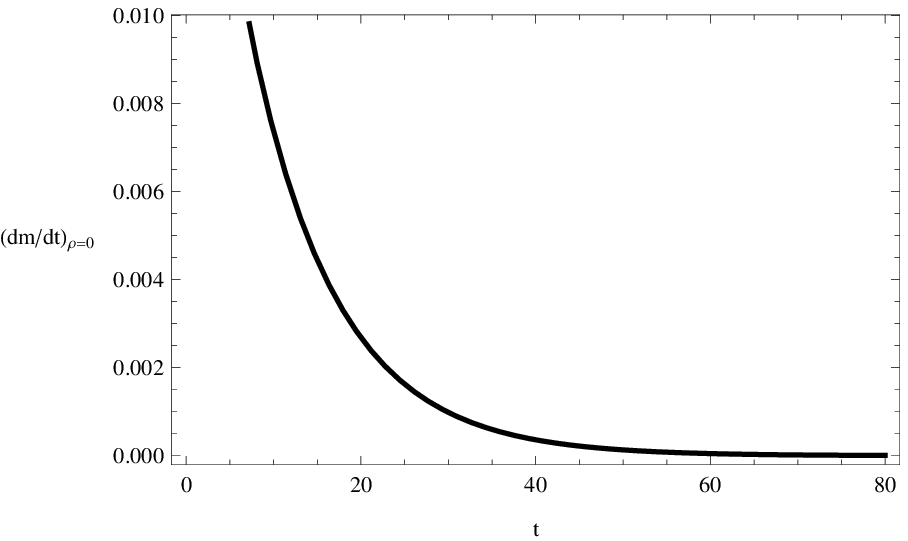} \caption{Temporal evolution of the mass flow rate across the stellar surface, in the absence of $\rho$} \label{fig4}
\end{figure}

\begin{figure}[t]
\centering
\includegraphics[scale=1.0]{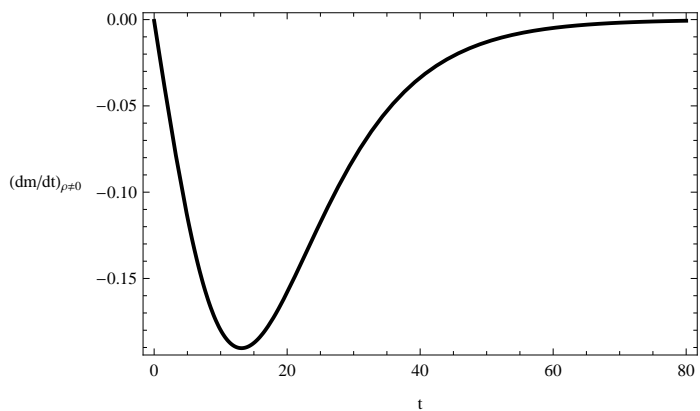} \caption{Temporal evolution of the mass flow rate across the stellar surface, in the presence of $\rho$} \label{fig4}
\end{figure}

\newpage
\subsection{Energy conditions with $\rho\neq0$}

In any radiating relativistic model in astrophysics, a test of the
strong, dominant and weak energy conditions are imperative in order
to establish whether the model is valid and physically reasonable. This has been demonstrated in the seminal study by Kolassis
\emph{et al} \cite{koletal} and in the recent treatment by Govender
\emph{et al} \cite{gov12}, for dense stellar objects with heat
conduction. The weak, dominant, and strong energy conditions are
described respectively, in the generalised Vaidya spacetime, by the following

\begin{subequations}
\label{encons}
\begin{eqnarray}
W=\mu-\left(qB-\rho\right)+\Delta \geq0,\label{weak}\\
\nonumber\\
D=\mu-3\left(qB-\rho\right)+\Delta \geq0,\label{dominant}\\
\nonumber\\
S=\left(qB-\rho\right)+\Delta \geq0, \label{strong}
\end{eqnarray}
\end{subequations}
where $\Delta$ is given by
\begin{equation}
\Delta=\sqrt{\left[\left(qB-\rho\right)+q\right]^2-4q^2}.\label{delta}
\end{equation}
In the limit when $\rho=0$ the classical definitions for the energy
conditions are regained.

The corresponding numerical profiles for
the parameters $W$, $D$, and $S$ are given in Fig. 13-15.
Fig. 13 demonstrates that the weak energy condition is satisfied
since the parameter $W$ is regular, continuous and non-negative for
all times. The dominant energy condition is also obeyed as seen from
the profile for $D$ in Fig. 14. However, the strong energy
condition is clearly violated since $S$ is always non-positive, in
Fig. 15. We must point out, though, that this result is not
entirely detrimental for the model since it has been established by
Kolassis \emph{et al} \cite{koletal}, that the strong energy
condition can be allowed to fail provided that the overall pressure
of the matter distribution is negative. This works perfectly well
with the generalised model with $m=m(v,r)$ since the pressure $P$ of the null fluid
is negative for a constant density (as highlighted earlier in
section 2.2) and in addition, the radiating fluid pressure is also
negative (from Fig. 3), at the surface of the star. We may then
conclude that the generalised null fluid model satisfies
collectively, all of the energy conditions and remains physically
acceptable.

\begin{figure}[t]
\centering
\includegraphics[scale=0.85]{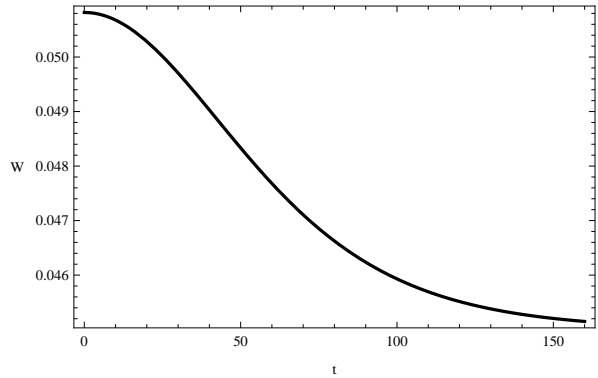} \caption{Temporal evolution of the weak energy condition parameter at the radiating surface, subject to the influence of the null fluid energy density $\rho$}\label{fig4}
\end{figure}

\begin{figure}[t]
\centering
\includegraphics[scale=0.85]{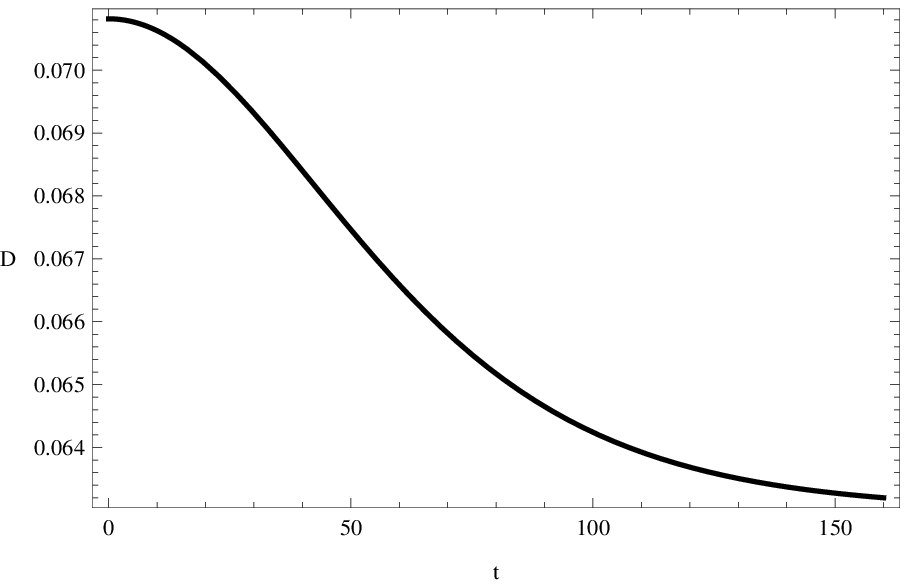} \caption{Temporal evolution of the dominant energy condition parameter at the radiating surface, subject to the influence of the null fluid energy density $\rho$}\label{fig4}
\end{figure}

\begin{figure}[t]
\centering
\includegraphics[scale=0.85]{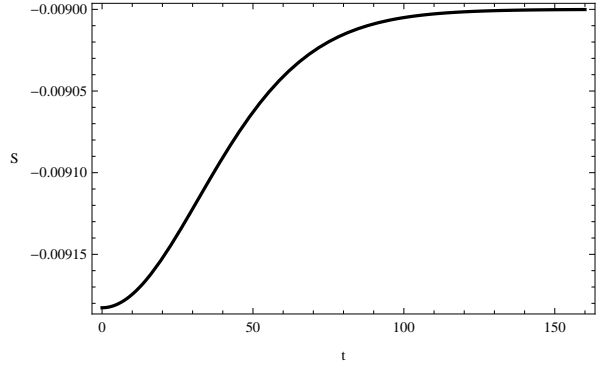} \caption{Temporal evolution of the strong energy condition parameter at the radiating surface, subject to the influence of the null fluid energy density $\rho$}\label{fig4}
\end{figure}

\subsection{The influence of the exterior null fluid on the nonadiabatic
gravitational collapse of the relativistic star}

A relativistic star can radiate away its heat energy in two
possible scenarios: in the process of nonadiabatic gravitational
collapse and when the object is in hydrostatic equilibrium undergoing thermal cooling. Our exact solution can be applied to the
former case in particular and it would be a point of interest to
determine what effect the exterior null fluid energy density $\rho$
could have on the collapse rate at the boundary.\\

The rate $\Theta$ of gravitational collapse for a
spherically symmetrical shear-free fluid with geodesic particles is given by

\begin{eqnarray}
\Theta &=& \frac{3\dot{B}}{AB},\label{collrate}
\end{eqnarray}
for a comoving 4-velocity $u^a=\delta^a_{0}$.
With the solution $(\ref{exactsolution})$, for
the situation where the additional null fluid is present on the
outside, the analytical form for the rate of collapse is
\begin{equation}
\Theta_{\rho\neq0}=\frac{3\left(b^2-r^2\right)\left(b^2d^2\rho-\alpha
^2\right)\sech\left[\frac{1}{2}t\sqrt{\frac{\alpha^2}{b^2d^2}-\rho}
\right]^2}{2bd\left[\left(b^2+r^2\right)\alpha+bd\left(b^2-r^2\right)\sqrt{\frac{\alpha^2}{b^2d^2}-\rho}\tanh
\left[\frac{1}{2}t\sqrt{\frac{\alpha^2}{b^2d^2}-\rho}
\right]\right]}.\label{gencrate}
\end{equation}
In the standard Vaidya model considered by Thirukkanesh and Maharaj
\cite{thiru&mah09}, with $\rho=0$, we have that
\begin{equation}
\Theta_{\rho=0}=\frac{3\alpha\left(b^2-r^2\right)\exp\left(\frac{\alpha(t+e)}{bd}\right)}{\left[r^2-\exp\left(\frac{\alpha(t+e)}{bd}\right)\right]
\left[b^2-\exp\left(\frac{\alpha(t+e)}{bd}\right)\right]}.\label{stdcrate}
\end{equation}
It is quite clear from equations $(\ref{gencrate})$ and
$(\ref{stdcrate})$ that in the general case $\Theta_{\rho\neq0}$ has a
strong dependence on the density $\rho$ and is far more involved
than in the standard case $\Theta_{\rho=0}$. The numerical profiles corresponding to
$(\ref{gencrate})$ and $(\ref{stdcrate})$ are given respectively by
Fig. 16 and 17.
\begin{figure}[t]
\centering
\includegraphics[scale=1.0]{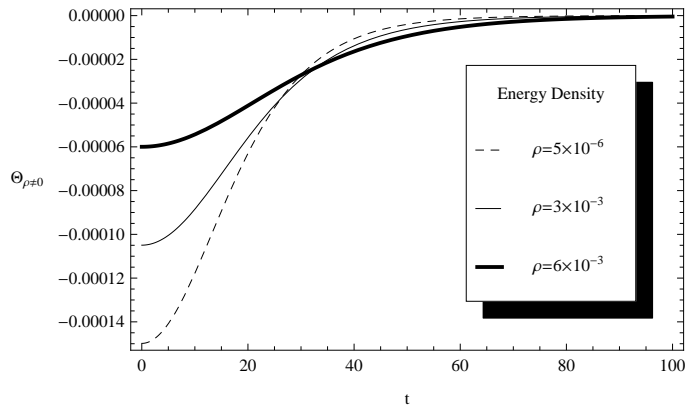} \caption{The collapse rate for varying null fluid energy density} \label{fig4}
\end{figure}

\begin{figure}[t]
\centering
\includegraphics[scale=1.0]{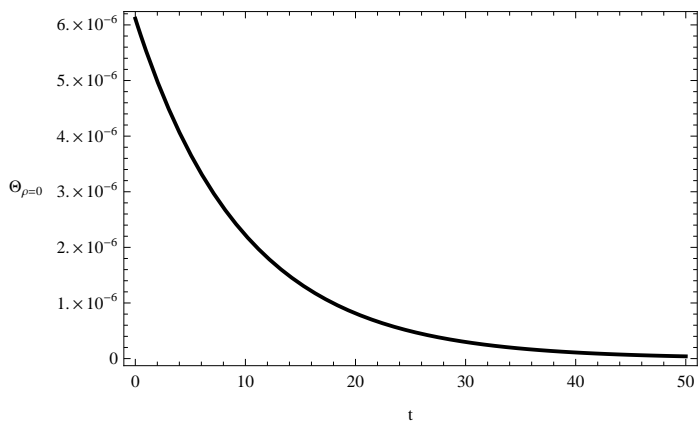} \caption{The collapse rate for $\rho=0$} \label{fig4}
\end{figure}
From Fig. 16 it is evident that the rate of collapse is
negative when the null fluid is present in the exterior of the
radiating body. In other words the non-vanishing energy density
$\rho$ actually slows down the collapse process until such time when
it reaches a complete halt and the relativistic star becomes stable
again. It can also be seen that as the magnitude of the density is
increased, $\Theta_{\rho\neq0}$ becomes less negative at earlier times
and eventually goes to zero at late times. The plot in Fig. 17
illustrates that for vanishing density, the rate of collapse is
always positive and indicates that the collapse itself is occurring
reasonably faster.

\newpage
\section{Discussion}
In this paper we modeled a spherically symmetric relativistic
radiating star that has vanishing shearing stresses and radial heat
flow in the interior while its local exterior region consists of a
two-fluid system comprising pressureless radiation and a general
null fluid. The Type \textrm{II} fluid possesses a non-zero pressure
and energy density and in particular we have investigated the
influence of the density $\rho$ on the physical behaviour of the
dissipating stellar fluid at the surface. We demonstrated that
the generalised junction condition $p=qB-\rho$ for a null fluid
scenario can be integrated as a nonlinear differential equation, at
the boundary and generated an exact solution in the special
case when the interior fluid particles are in geodesic motion and
the exterior density is constant. Our solution, although similar in
structure to that obtained by Govender and Thirukkanesh
\cite{gov&thiru09}, has a very different interpretation, depending
strongly on the magnitude of $\rho$ and can be cast in a more
realistic astrophysical context. The generalised two-fluid model
with our exact solution has been compared to the standard model for
pure radiation, studied by Thirukkanesh and Maharaj \cite{thiru&mah09},
by constructing numerical profiles for various physical
parameters associated with the stellar fluid at the boundary.

We established that the constant null fluid energy density has a
marked impact and significantly reduces the pressure $p_{\Sigma}$
and luminosity $L_{\Sigma}$ when compared to the limiting case when
$\rho=0$, where the pressure and luminosity are positive, finite and
consistent with well established results for extremely dense stellar
objects in the strong gravity regime. The flow of mass across the
radiating surface was also probed and it has been demonstrated that
the density $\rho$ actually enhances the outgoing flow; this is
suggestive that the null fluid on the outside could more likely be
interpreted as a distribution of particles that are sourced from
nonadiabatic and particle production processes originating within
and at the surface of the star. This is in contrast to
other treatments where the null fluid has been considered to be
an ambient string system in four dimensions. In light of this our
model may offer a more realistic description for dense stars
in a relativistic setting. By utilising the interior matter
variables and the exact solution $(\ref{exactsolution})$ we also tested the strong, dominant and weak energy conditions when the
constant density null fluid is prevalent. Numerical profiles
indicate that the weak and dominant conditions are satisfied while
the strong condition is violated. However this does not severely
affect the validity of the model in any way as the overall pressure is negative. Since radiating
models are crucial for studying the nonadiabatic gravitational
collapse of stellar objects, the rate of collapse, for both
situations has also been examined. It has been found that $\rho$
slows down the collapse process while in the case of null radiation only
it occurs faster. Our model can be further improved on by including the effects of shear
and bulk viscosity and analysing the dynamical stability of the
stellar matter configuration. Furthermore, if the null fluid is to
be taken as a system of particles like neutrinos for example then
the model can be appropriately adapted by incorporating
non-vanishing neutrino fluxes in the interior. However, these
considerations would consequently render the model more complicated
and in particular the junction condition increasingly difficult to
solve, even with the robust mathematical techniques that are
currently available. These endeavors will be the subject of an ongoing investigation.\\

\begin{center}
\large{\bf Acknowledgements}
\end{center}
\vspace{0.5cm} \noindent GG, BPB and SDM thank the National Research
Foundation and the University of KwaZulu-Natal for financial
support. SDM acknowledges that this work is based upon research
supported by the South African Research Chair Initiative of the
Department of Science and Technology and the National Research
Foundation. We thank S{\o}ren Greenwood for his assistance
with computing software related issues. The authors would also like
to thank Dr. Megandhren Govender and Dr. Rituparno Goswami for constructive
criticisms and useful discussions.

\end{document}